\input{aipcheck}

\documentclass[
    ,final            
  ]
  {aipproc}

\layoutstyle{8x11single}

\def\aj{\emph{AJ}}
\def\apj{\emph{ApJ.}}
\def\apjl{\emph{ApJ. Lett.}}
\def\aap{\emph{A.\& A.}}

\def\mnras{\emph{MNRAS}}
\def\nat{\emph{Nature}}

\def\pasp{\emph{PASP}}

\begin{document}

\title{Follow the BAT: Monitoring $Swift$ BAT FoV for Prompt Optical Emission from Gamma-ray Bursts}





\author{T. N. Ukwatta}{
  address={Michigan State University, Physics and Astronomy Department, East Lansing, MI 48824, USA}
}

\author{J. Linnemann}{
  address={Michigan State University, Physics and Astronomy Department, East Lansing, MI 48824, USA}
}

\author{K. S. Dhuga}{
  address={The George Washington University, Washington, D.C. 20052, USA}
}

\author{N. Gehrels}{
  address={NASA Goddard Space Flight Center, Greenbelt, MD 20771, USA}
}

\begin{abstract}
We investigate the feasibility of implementing a system called
`Follow the BAT' that will coordinate ground-based robotic optical
and near infrared (NIR) telescopes to monitor the Swift BAT
field-of-view (FoV). The system will optimize the monitoring
locations in the BAT FoV based on individual robotic telescopes'
location, FoV, sensitivity and local weather conditions. The aim is to perform coordinated BAT FoV monitoring by professional as well as amateur astronomers
around the world. The scientific goal of the proposed system is to
facilitate detection of prompt optical and NIR emission from GRBs,
especially from short duration GRBs. We have performed a Monte
Carlo simulation to investigate the feasibility of the project.
\end{abstract}

\maketitle


\section{Introduction}

Our understanding of Gamma-ray Bursts (GRBs) progressed very
rapidly after the detection of multi-wavelength afterglows.
Usually, a favorably positioned GRB gets fairly good
multi-wavelength afterglow coverage. However, multi-wavelength
observations of the prompt emission are sparse. The reason is
obvious. The location and the time of a GRB cannot be predicted,
and currently, it is not possible to monitor the whole sky using
multi-wavelength instruments. Despite this fact, multi-wavelength
observations of prompt emission of GRBs are crucial to enhance our
understanding of GRB physics.


Initial detection of prompt optical emission from GRB 990123
\citep{Akerlof1999} has been followed by few bursts that are
observed in optical wavelengths during the gamma-ray emission.
Some of these detections are due to fortunate simultaneous
observations from multiple detectors (e.g. GRB 080319B;
\citep{Racusin2008}) or when the GRB was preceded by a strong
precursor event (e.g. GRB 061121; \citep{Page2007}). In addition,
a number of fast slewing robotic optical telescopes such as
ROTSEIII \citep{Akerlof2003}, TAROT \citep{Klotz2009}, REM
\citep{Zerbi2001}, and RAPTOR \citep{Vestrand2004} have been used
to catch prompt optical emission while the gamma-ray emission is
still ongoing. However, all these fast slewing telescopes are at
the mercy of the propagation time of the burst notice from the
space craft to the telescope and also on the slewing time of the
telescope. There has never been a systematic coordinated effort to
detect prompt optical emission from GRBs independent of the notice
propagation time and the slewing time.

Even though a few cases of prompt optical detections have been
reported involving long duration bursts, so far there has been no
detection of prompt optical emission from short duration bursts.
If the progenitors of short duration bursts are merging compact
objects, then there are theoretical predictions of possible early
optical/UV emission from an accompanying `mini-supernove'
\citep{Li1998}, caused by the ejection of radioactive material
from the merging system. This optical signal fades away very
quickly and can only be detected by monitoring a nearby short
burst location at the burst onset.

Based on the current sample of bursts with prompt optical
coverage, it seems there are two categories of GRBs. The first
category is characterized by bright optical transients that are
uncorrelated with the gamma-ray emission (e.g. GRB 990123;
\citep{Akerlof1999}). The second category has weak optical
emission correlated with the gamma-ray emission (e.g. GRB 041219A;
\citep{Vestrand2005}). The optical emission of the first category
is thought to originate from reverse shocks \citep{Jin2007}
whereas for the second category optical emission is generally
interpreted as the low-energy tail from the gamma-ray emission
\citep{Genet2007}. However, interpretation of the second category
is questionable after detection of bright correlated optical
emission from GRB 080319B.

Another interesting question is whether GRB precursors are
associated with any optical or NIR emission. Currently there are
no observational data on precursors other than on gamma-ray
wavelengths. Evidence for precursors has been reported both in
short \citep{Troja2010} and long \citep{Koshut1995,Lazzati2005}
duration bursts. There are number of interpretations for GRB
precursors such as 1) photospheric blackbody emission just before
the transition to optically thin regime, 2) due to interaction of
the relativistic jet with the stellar envelope, or 3) due to
two-step collapse of the central engine. Detection of optical
emission coincident with a gamma-ray precursor event may cause us
to rethink these interpretations.

Currently, there are robotic telescopes that can slew to the
location of the burst as fast as $\sim$5 seconds after the receipt
of the GCN notice \citep{Barthelmy1998}. However, the time
difference between the trigger time and the arrival time of the
GCN notice is significant. Typically this delay can be $\sim$ 15
seconds. If one wants to observe prompt optical emission from
bursts which last for less than $\sim$ 20 seconds, especially
short bursts, then one way to observe them is by  monitoring the
BAT FoV.

\section{Proposed Observing Program}

Inspired by successful `citizen science' projects such as Galaxy
Zoo and NASA's Stardust@Home projects, we propose to implement a
system called `Follow the BAT' to facilitate monitoring of the BAT
FoV for prompt optical and NIR emission from GRBs. This observing
program could be specially aimed at amateur astronomers around the
world. Proliferation of amateur robotic telescopes with high
quality CCD cameras have opened a new avenue to study prompt
optical emission from GRBs. The basic objective of the `Follow the
BAT' system is to coordinate large number of ground based robotic
telescopes to monitor different patches of the BAT FoV. Selection
of various patches in the BAT FoV will be done based on the number
of available telescopes, individual robotic telescopes' location,
FoV, sensitivity and local weather conditions.

Some of the important observational questions that can be
potentially answered by such an observing program are given below:
\begin{itemize}
    \item Is there any prompt optical emission from short duration GRBs?
    \item Are GRB precursors associated with any simultaneous optical emission?
    \item What fraction of GRBs have prompt optical emission at the trigger time?
    \item How is the prompt optical light curve correlated with the gamma-ray light curve? Are there two populations?
    \item What are the variability time scales of the GRB prompt optical and NIR emission?
\end{itemize}

The proposed system is envisioned to have a web interface which
will allow people to register their robotic telescopes and obtain
useful information such as 1) the BAT pointing direction in
real-time, 2) all participating robotic telescopes, 3) regions of
the BAT FoV allocated to specific robotic telescope, and 4)
covered and uncovered regions of the BAT FoV at a given time.

\begin{figure}[htp]
\centering
\includegraphics[width=0.5\textwidth]{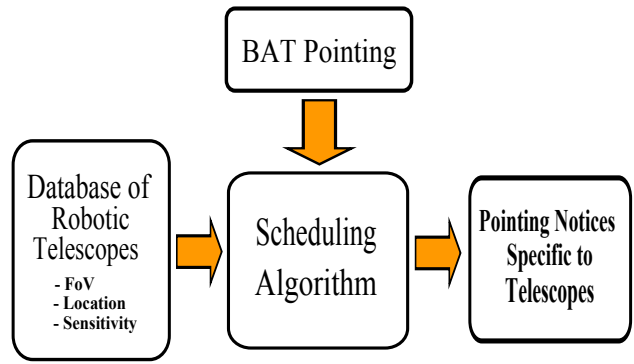}
\caption{Schematic block diagram of the proposed `Follow the BAT'
software system. }\label{diagram}
\end{figure}

Unlike the GCN system \citep{Barthelmy1998} which sends notices to
large number of recipients, the proposed system will send
customized targeted messages to individual registered robotic
telescopes. The individual messages will be sent via email or
socket connections. These customized messages will have assigned
pointing locations for the robotic telescope. Another advantage
for the telescopes that are pointing inside the BAT FoV is that
even when they do not necessarily catch the burst they are
relatively close by to the burst location and would therefore be
able to slew to it relatively quickly. Hence, by participating in
the `Follow the BAT' program, robotic telescopes can potentially
increase their response times for burst triggers.

A schematic block diagram of the proposed `Follow the BAT'
software system is shown in the Figure~\ref{diagram}. The
scheduling algorithm will check the BAT FoV observability of each
robotic telescope and assign them to different parts of the BAT
FoV. In doing this the algorithm will consider number of parameters
such as individual telescopes' FoV and sensitivity. In addition, it will also
consider the bore-sight angle distribution of previously detected
$Swift$ bursts.


\section{Feasibility and Challenges}

\begin{figure}[htp]
\centering
\includegraphics[width=0.45\textwidth]{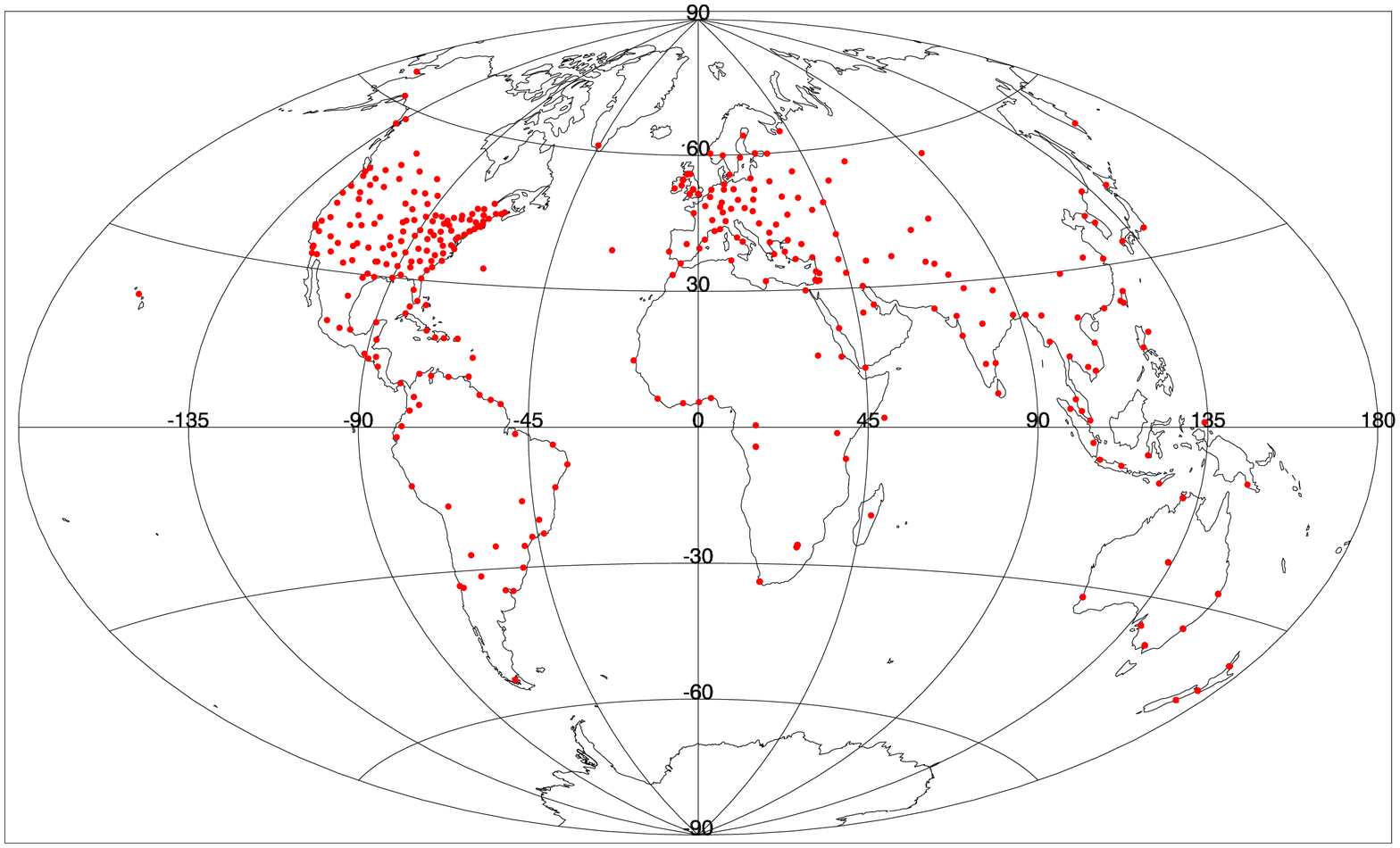}
\includegraphics[width=0.45\textwidth]{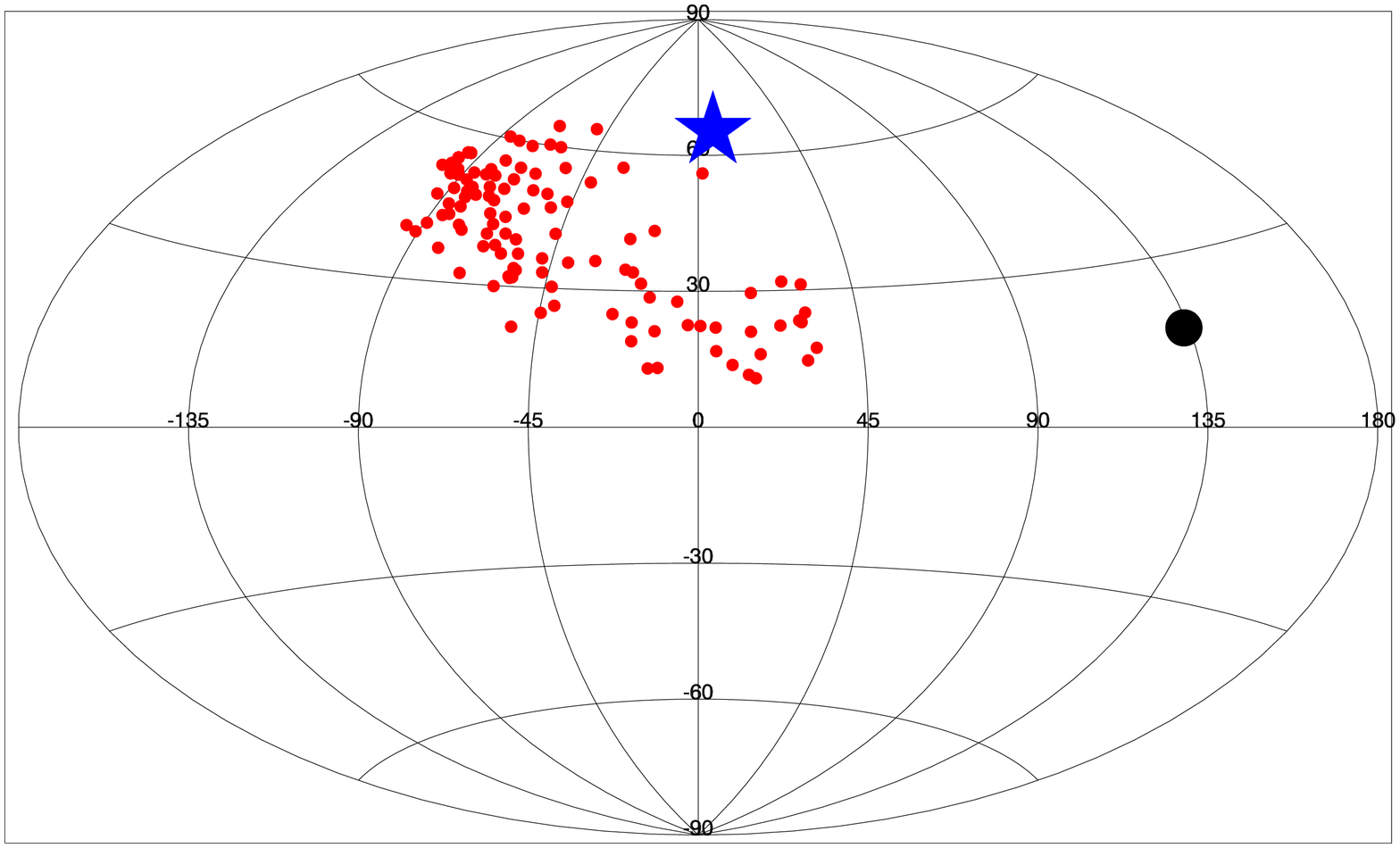}
\caption{Left Panel: Aitoff projection of the world map showing
the distribution of telescopes used in the simulation. Right
Panel: Galactic Aitoff projection of the location of the burst
(blue star), the Sun (black filled circle) and zenith coordinates
of telescopes that can observe the burst (red circles).
}\label{projections}
\end{figure}

In order to investigate the feasibility of the project, we
performed a Monte Carlo simulation to study the probability of
detection of prompt optical emission from $Swift$ GRBs. We assumed
that $Swift$ detects about 100 GRBs per year distributed
isotropically in the sky and also throughout the year. We have
distributed robotic telescopes in such a way that they roughly
trace the major cities in the world as shown in the left panel of
Figure~\ref{projections}. Then for each burst we tracked the path
of the Sun and selected a set of telescopes away from the Sun and
within few hours from the burst location to monitor the BAT FoV.
The right panel of Figure~\ref{projections} shows the selected set
in red circles for a given burst. We calculated the probability of
detecting a given burst by summing FoV solid angles of all
telescopes in the selected set and then dividing by the BAT FoV
solid angle ($60.0^{0} \times 60.0^{0}$). We repeated this
procedure for every simulated burst, while changing the number of
total telescopes available and FoV of telescopes.

\begin{figure}[htp]
\centering
\includegraphics[width=0.46\textwidth]{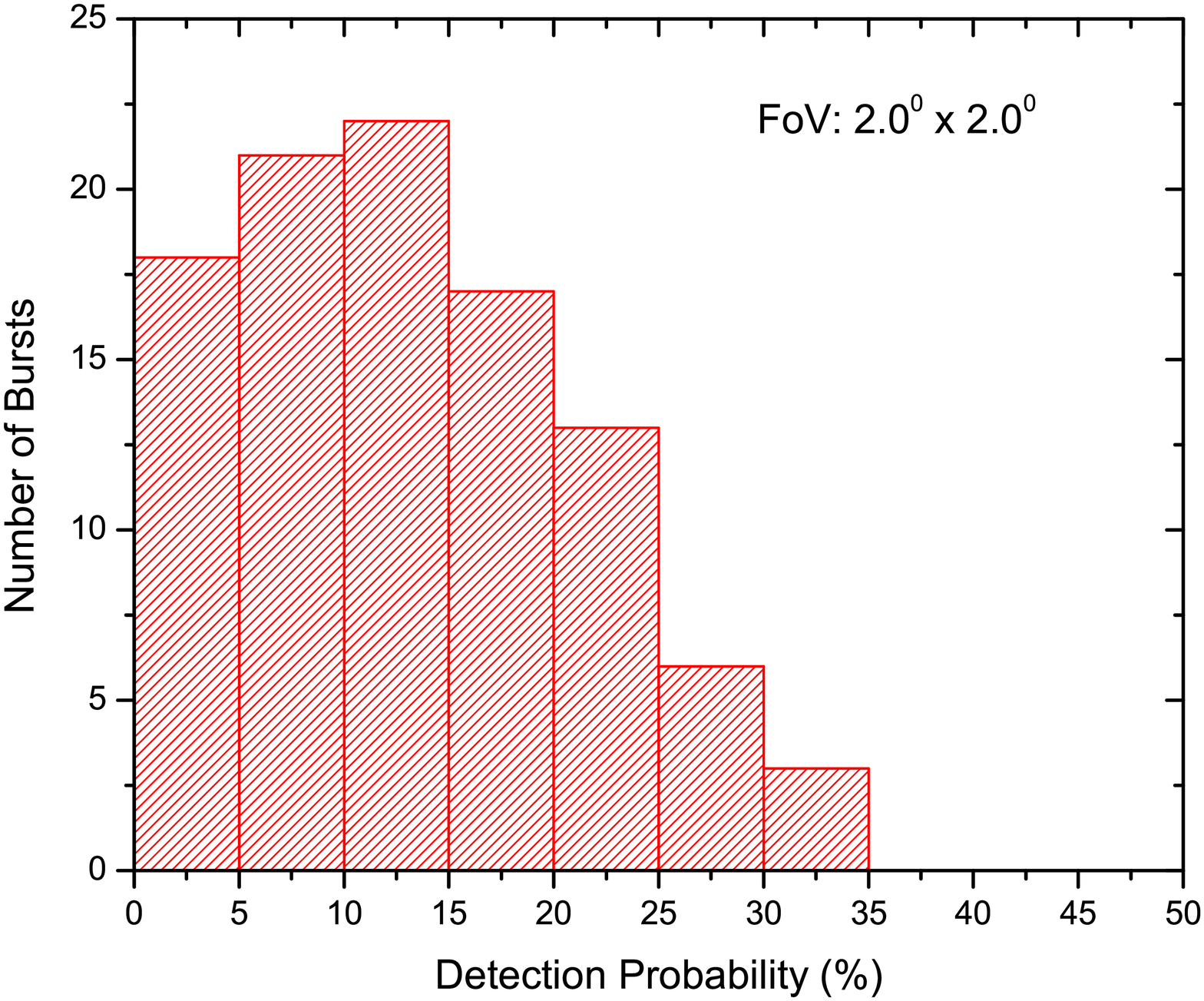}
\includegraphics[width=0.45\textwidth]{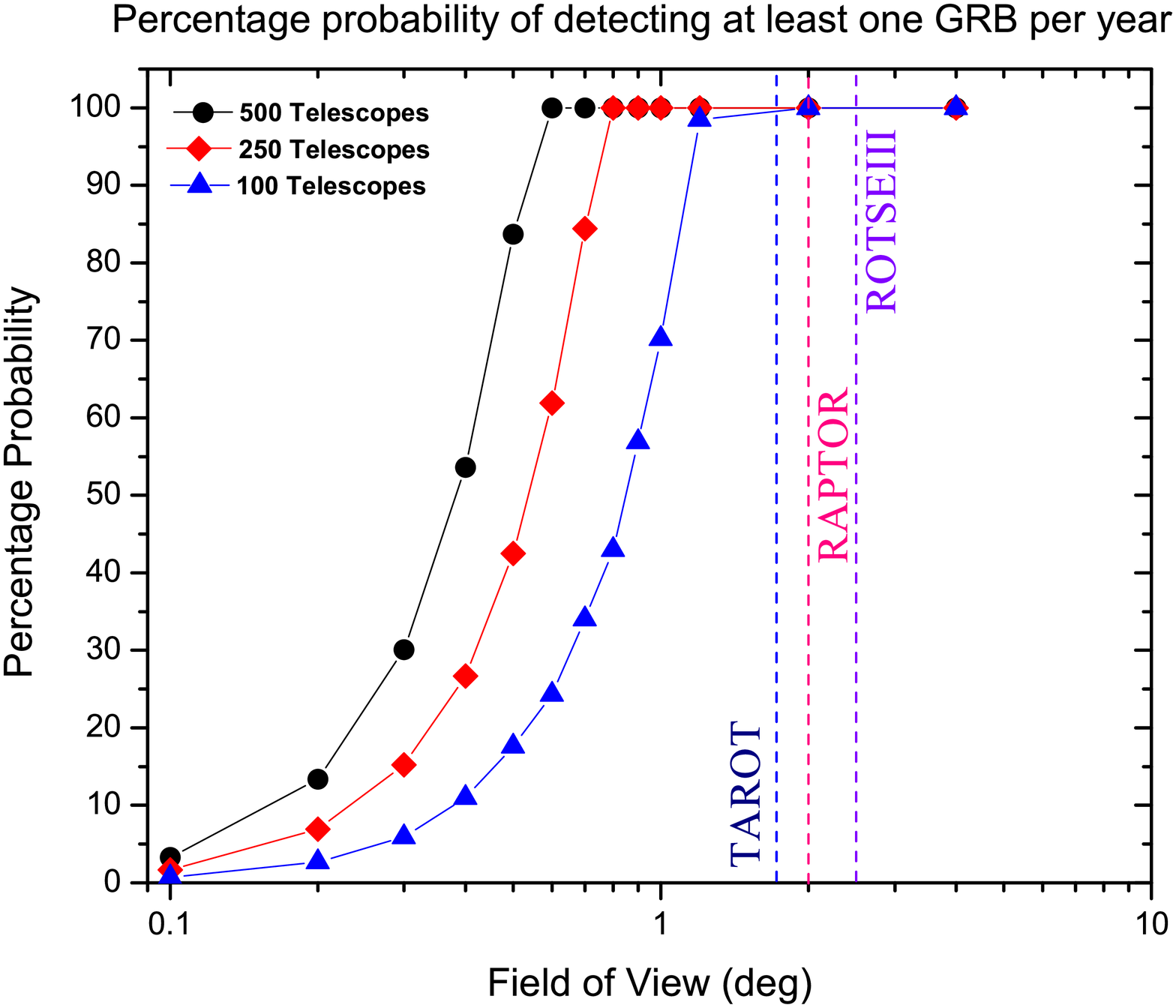}
\caption{Left Panel: Histogram of detection probability of 100
bursts with 500 telescopes with FoV of $2.0^{0} \times 2.0^{0}$.
Right Panel: The percentage probability of detecting at least one
burst per year as a function of the FoV when 100, 250, and 500
telescopes are participating in the program.}\label{prob_histo}
\end{figure}

The results of our simulation are shown in
Figure~\ref{prob_histo}. For these particular simulations we have
assumed all the robotic telescopes have the same FoV. The number
of participating telescopes in the observing program are taken to
be 100, 250 or 500.
The left panel of Figure~\ref{prob_histo} shows the distribution
of percentage probability values of detecting bursts with 500
telescopes with FoV of $2.0^{0} \times 2.0^{0}$.

The right panel of Figure~\ref{prob_histo} depicts the percentage
probability of detecting at least one burst per year as a function
of the FoV. With 500 telescopes participating, it is possible to
detect at least one burst using robotic telescopes with FoV of
$0.6^{0} \times 0.6^{0}$. This value is close to the value of a
typical FoV of an amateur robotic telescope. It is also
interesting to note that if we have about 100 telescopes
participating with FoV of $1.0^{0} \times 1.0^{0}$, we can detect
at least one burst per year. The simulation did not take into
account the moon fraction and fraction of the time BAT FoV is
covered by clouds.

In practice, however, there are number of serious challenges to
carry out this type of observing program:
\begin{itemize}
    \item Number of telescopes needed to perform the
    observation is relatively large.
    \item Number of telescopes required can be brought down by
    increasing the FoV. However, the increment of the FoV may
    result in loss of sensitivity.
    \item Currently, the number of amateur robotic observatories
    capable of monitoring BAT
    FoV is small. However, this number is growing steadily.
    \item Professional observatories which are capable of
    monitoring the BAT FoV have dense observation schedules and
    are unlikely to participate in the program.
\end{itemize}

Despite these difficulties, `Follow the BAT' type of observing
program could enable the investigation of the optical properties
of the prompt emission of GRBs, particularly in short ones. It
could also enable the study of the precursors in optical
wavelengths for the first time. In addition, the project will
potentially allow and attract the participation of amateur
astronomers and their telescopes around the world.

\section{Discussion and Conclusion}

In this paper we have investigated the feasibility of implementing
an observation program to monitor the $Swift$ BAT FoV for prompt
optical and NIR emission. The simulation results show, even though
there is a fair chance of detecting a few GRBs with prompt optical
emission, the amount of effort needed to carry out the task is
substantial. However, astonishing $\sim$ 5th magnitude prompt
optical emission from GRB 030319B and the detection of 12
magnitude optical flare from very high redshift burst, GRB 050904
(with redshift of 6.29) indicate that we should be able detect
optical emission from GRBs up to very high redshift with varying
degrees of brightness. With moderate exposure times ($\sim$
seconds) amateur telescopes may be able to reach these magnitude
levels and observe prompt optical emission from GRBs. The decades
old observation model -- space-based gamma-ray observatories
discovering bursts and ground-based observers following -- is
approaching its limits after the highly successful $Swift$
mission. Now may be the time to change this traditional
observation model and try to observe GRBs in multi-wavelengths as
they happen.



\bibliographystyle{aipproc}   


\IfFileExists{\jobname.bbl}{}
 {\typeout{}
  \typeout{******************************************}
  \typeout{** Please run "bibtex \jobname" to optain}
  \typeout{** the bibliography and then re-run LaTeX}
  \typeout{** twice to fix the references!}
  \typeout{******************************************}
  \typeout{}
 }

\end{document}